# Stress relaxation and creep experiments with the atomic force microscope: a unified method to calculate elastic moduli and viscosities of biomaterials (and cells)

**Susana Moreno-Flores[1], Rafael Benitez[2], María dM Vivanco[3], and José Luis Toca-Herrera[1]**

[1] Biosurfaces unit, CIC BiomaGUNE, Paseo Miramón 182, 20009 San Sebastián-Donostia, Spain

[2] Dept. Mathematics, Centro Universitario de Plasencia, Universidad de Extremadura, Avda. Virgen del Puerto 2, 10600 Plasencia, Spain

[3] Cell Biology & Stem Cells Unit, CIC BioGUNE, Parque tecnológico de Bizkaia, Ed. 801A, 48160 Derio, Spain

* Corresponding author:
José L. Toca-Herrera
CIC BiomaGUNE
Paseo Miramón, 182
20009 San Sebastián (Spain)
Tel: ++34 943 00 53 13
Fax: ++34 943 00 53 01
E-mail: jltocaherrera@cicbiomagune.es

## INTRODUCTION

Physical forces influence cell's development, cell's function and cell's fate but they also can induce pathological transformations (Butcher et al, 2009). Among other techniques (Hochmuth, 2000; Thoumine and Ott, 1997; Yamada et al; Valberg and Albertini, 1985), atomic force microscopy has been used to mechanically characterize adherent living cells (Kuznetsova et al, 2007). The non-rheological (Alcaraz et al, 2003; Mahaffy et al, 2004) approach consists in applying a local normal load on the cell surface. The load is exerted by a micromechanical cantilever that is brought to contact the biomaterial's surface through the cantilever tip, usually a pyramid, a cone or a bead attached to its free end. The force is monitored together with the sample's indentation, and the resulting force-distance curve is analysed in terms of elastic contact mechanics obtaining the Young's moduli of the material. The model of Sneddon, which takes into account a conical tip impinging on an elastic semi-infinite space (Sneddon, 1965), together with the Hertz model for spherical tips (Hertz, 1885) are mostly used. In this approach the tip-cell adhesion is assumed to be negligible and the experiments are made under deformations not larger than 10 % (Dimitriadis et al, 2002).

However, the models do not take into account, for example, the fact that on living cells (and other viscoelastic biomaterials) the force-indentation curves may vary with the loading rate, the indentation or the applied load (Bremmell et al, 2006; Li et al, 2008; Zhao, et al, 2006). Thus analysis according to an elastic contact mechanics model would lead to Young's moduli that may depend on loading rate and applied load; therefore a new approach might be needed to characterised the mechanical properties of biomaterials.

In this short paper we present a (theoretical) unified method to obtain the viscoelastic parameters on living cells or other biomaterials from stress relaxation and creep compliance experiments using the atomic force microscope. This methodology, together with the already reported STREM (Moreno-Flores et al, 2010), aims to define an unified experimental framework to address the heterogeneity and viscoelasticity of biomaterials through scanning probe microscopies.

## GENERAL METHODOLOGY

*Force & height-time curves*. Measurements were carried out on different breast cancer cells (MCF-7) with a Nanowizard II (JPK Instruments, Germany) coupled with a transmission optical microscope (Axio Observer D1 Zeiss, Germany). Uncoated SiN cantilevers of nominal spring constant of 0.06 N/m (MLCT, Veeco Instr., USA) can be used. The cantilevers were be cleaned in acetone and ethanol to remove impurities and their spring constants evaluated by the thermal method.

Individual force-time curves were be recorded on different parts of the sample at different speeds and loads. The AFM was kept in contact with the cell surface (e.g. for 2 seconds) at constant force and at constant height to obtain the stress relaxation and creep curves, respectively.

*Deformation*. Material´s deformation ($\Delta l_0$) in experiments at constant height can be calculated from displacement-time curves at different loads ($L_i$) (Melzak et al, 2010). The constant height attained by the piezo-element at each load L ($Z_L$) is registered and subtracted from the height reached at the lowest load $L_0$. The resulting quantity ($Z_L$-$Z_{L0}$) is thus a relative displacement, which is influenced by the applied load, the sample deformation and the cantilever deflection ($Z_L$-$Z_{L0}$ = $\Delta l_0^{(rel)}+\delta_{cant}$). The relative sample deformation at load $L_i$ is obtained by subtracting the relative cantilever deflection ($\delta_{cant}$ = $(L_i-L_0)/k$) to the relative piezo-element displacement. The absolute sample deformation at each load can be obtained by adding an offset, $\Delta l_{offset}$ ($\Delta l_0 = \Delta l_{offset}+\Delta l_0^{(rel)}$). This offset is obtained by fitting the relative displacements to a power law of the form $\Delta l_0^{(rel)}$ = $\Delta l_{offset}+K\cdot L^{0.5}$. This approach assumes that the force is directly proportional to the square of the deformation, which agrees with the mechanistic model of the elastic half-space being impinged by a conical tip (Sneddon, 1965), and the reported efforts in assessing the mechanical properties of living cells as such (Radmacher, 1996).

## GETTING MECHANICAL PARAMETERS (RESULTS)

The atomic force microscope permits to monitor the time evolution of the cantilever's force and the cantilever's vertical position at all stages. When the AFM tip is brought into contact with the sample's surface for a definite time period, two type of experiments are possible (figure 1). In the so-called constant height mode (figure 1a), the cantilever's vertical position is set constant while the cantilever's force varies with time. In the constant force mode (figure 1b) the vertical position of the cantilever changes with time while the cantilever's force is kept constant. The first case deals with force relaxation tests while the second case resembles creep compliance tests. The variations of the vertical position of the cantilever can be directly related to the deformation of the biomaterial at constant force ($\Delta l(t)$).

Both force and deformation are proportional to the stress and strain, respectively, through the contact area and the cell's thickness (figure 1).

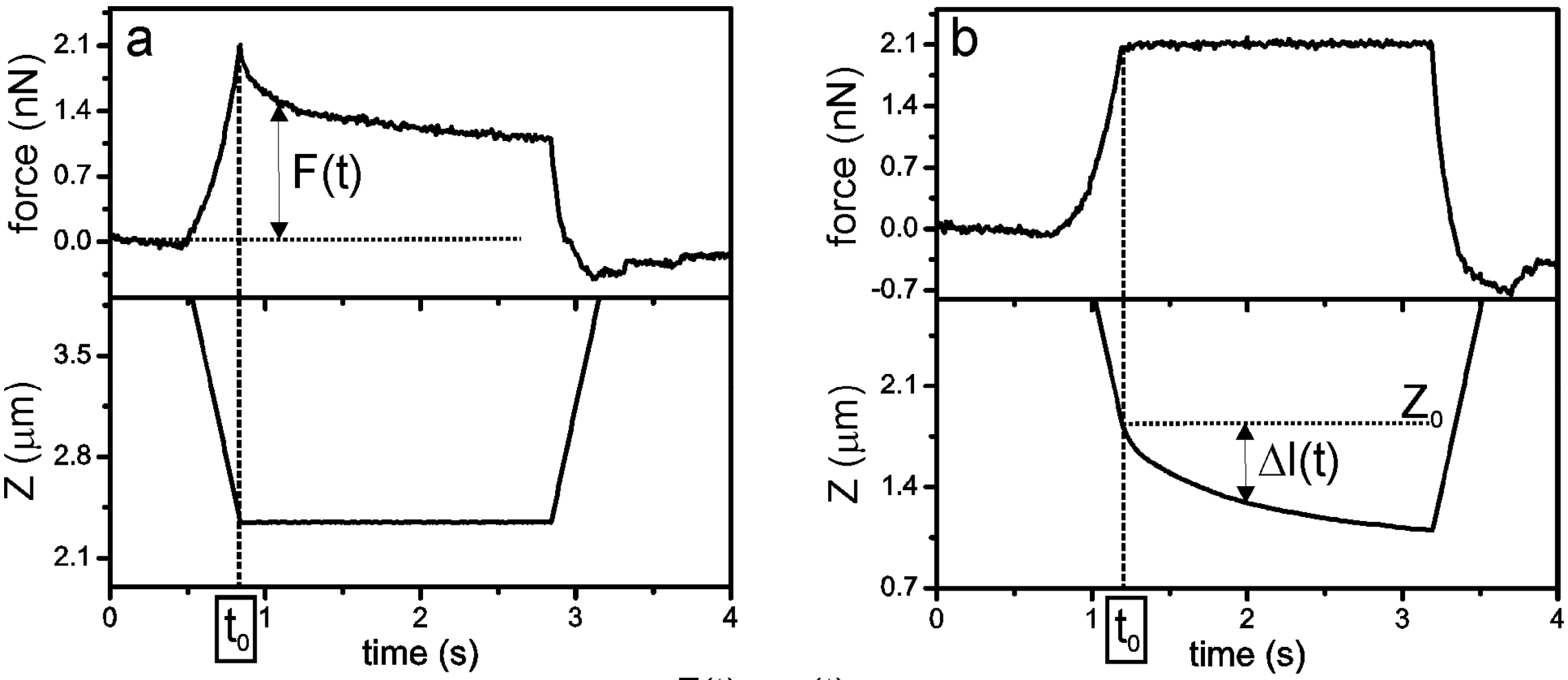

$$F(t) = \sigma(t) \times a_c$$
$$Z(t) - Z_0 = \Delta l(t) = \varepsilon(t) \times l_0$$

**Figure 1.** Example of force relaxation (a) and creep (b) experiments on the nuclear region of a living MCF-7 cell[1]. During the time the AFM tip is in contact with the cell (starting at $t_0$) either the force is set to vary with time (constant height mode), or the vertical position (constant height mode). Force (F) and deformation ($\Delta$l) are related to the stress and strain through the contact area ($a_c$) and the cell height ($l_0$).

As figure 1 shows the force in stress relaxation experiments decays with time while the cell's deformation increases with time. Both decays are related and should be analysed according to a common mechanistic model. The model has been previously proved successful in describing force relaxation experiments on the same type of cells (Moreno-Flores et al, 2010) and consists of a parallel arrangement of two Maxwell elements and a spring. In general terms, if the cell is subjected to constant deformation (assuming that the contact area does not change with time) the force decays bi-exponentially:

[1]MCF-7 cells were grown at 37ºC and 5% $CO_2$ in Dulbecco's modified eagle medium (DMEM, Sigma) supplemented with 8% fetal bovine serum (FBS, Sigma), 2% 200 mM L-glutamine, 0.4% penicilline/streptomicine (PEN/STREP, Sigma). For force measurements, the cells were subcultured on borosilicate glass coverslips (diameter 24 mm and 0.16 thickness) for one day. Prior to force measurements, the cells were washed in $CO_2$-independent cell medium (Leibowitz medium, L15, Sigma) and measured in the same medium at 37ºC.

$$F(t) = a_0 + a_1 \exp\left(\frac{-(t-t_0)}{\tau_1}\right) + a_2 \exp\left(\frac{-(t-t_0)}{\tau_2}\right) \qquad (1)$$

with $\tau_1 = \eta_1/E_1$ and $\tau_2 = \eta_2/E_2$ , being $E_1$ and $E_2$ the compressive moduli and $\eta_1$, $\eta_2$ the viscosities of the first and the second Maxwell element, respectively. Likewise, if a sudden and constant force is applied on the biomaterial´s surface, and the deformation follows a bi-exponential decay, we obtain:

$$Z - Z_0 = c_0 + c_1 \exp(x_1 t) + c_2 \exp(x_2 t) \qquad (2)$$

With $x_1$ and $x_2$ being both negative and non-linear functions of the compressive moduli and viscosities (the whole derivation is available in Appendix A):

$$\begin{gathered} x_1 = \frac{-r_1 + \sqrt{r_1^2 - 4r_0 r_2}}{2r_2}, x_2 = \frac{-r_1 - \sqrt{r_1^2 - 4r_0 r_2}}{2r_2} \\ r_0 = \frac{E_0 E_1 E_2}{\eta_1 \eta_2},\ r_1 = \frac{E_1}{\eta_1}(E_0 + E_2) + \frac{E_2}{\eta_2}(E_0 + E_1),\ r_2 = E_0 + E_1 + E_2 \end{gathered} \qquad (3)$$

Figure 2 shows a set of force relaxation curves obtained at different loads (black) and their respective fittings to a bi-exponential decay (red)[2]. The correlation coefficients range from 0.959 to 0.999 in both experiments. We should point out that a prerequisite to obtain a unique set of non-load dependent compressive moduli and viscosities is their constant behaviour within the applied load range.

[2] The measurements were performed on the apical zone of a living MCF-7 cell.

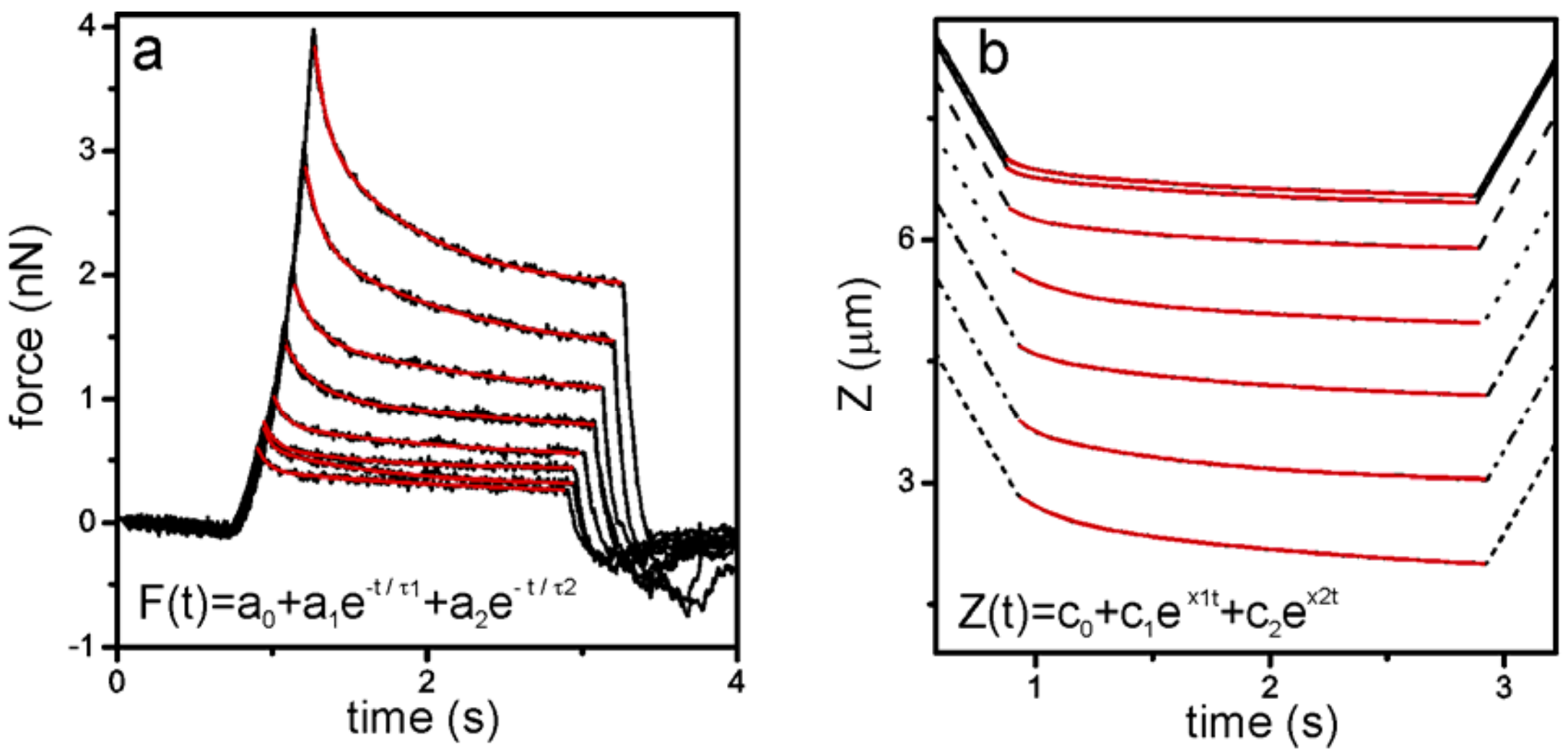

**Figure 2.** Example of force relaxation (a) and creep (b) curves obtained at different loads taken on the nuclear region of a MCF-7 cell (the curves in figure 2b are vertically displaced for clarity). The black lines are the experimental data and the red lines are the corresponding fits to the Zener's model.

## DISCUSSION

In order to obtain $E_0$, $E_1$, $E_2$, $\eta_1$ and $\eta_2$ we assume that the biomaterial´s compression is quick enough and that the shape of the perturbation does not influence the response of the biomaterial, while the contact area in the force relaxation experiments does not change with time. According to that, the proposed model predicts the following expressions for the compressive moduli and viscosities (see appendix 1 for derivation):

$$\begin{aligned}
E_0 &= \frac{A_0}{\varepsilon_0} = \frac{a_0 l_0}{\Delta l_0 a_c} \\
E_1 &= \frac{a_0 l_0}{\Delta l_0 a_c}\left[\frac{1}{\left(1-\frac{\tau_1}{\tau_2}\right)}\cdot\left(1+\frac{1}{x_1\tau_2}+\frac{1}{x_2\tau_2}+\frac{1}{x_1 x_2 \tau_2^2}\right)+\frac{1}{x_1 x_2 \tau_1 \tau_2}-1\right] \\
E_2 &= \frac{a_0 l_0}{\Delta l_0 a_c}\cdot\frac{1}{\left(1-\frac{\tau_1}{\tau_2}\right)}\cdot\left(-1-\frac{1}{x_1\tau_2}-\frac{1}{x_2\tau_2}-\frac{1}{x_1 x_2 \tau_2^2}\right) \qquad (4) \\
\eta_1 &= E_1\tau_1 \\
\eta_2 &= E_2\tau_2
\end{aligned}$$

where $a_c$ is the projection of the actual contact area along the normal direction (i.e., the direction along which the force is applied and measured), $\Delta l_0$ is the deformation of the biomaterial in stress relaxation experiments (which is assumed to be constant with time) and $l_0$ is the biomaterial´s height. Values for $a_c$ cannot be directly obtained from AFM experiments and assumptions are frequently made that are valid for perfectly elastic samples and small indentations (Hertz model and derived approaches, (Kuznetsova et al, 2007). If we assume that $a_c$ is the projected area defined by the surface of the biomaterial being in contact with the pyramidal tip (Mathur et al, 2001), the deformed surface follows the tip's shape and it extends along the normal direction. In this case, the projected area is $a_c = 2\Delta l_0^2 \tan^2 \alpha$, where $\alpha$ is the half-opening angle of the tip. Since the tip is a truncated pyramid with a sphere-like apex, the expression is valid as long as the deformation is higher than R(1-sin $\alpha$), where R is the curvature radius of the tip apex.

## CONCLUSION

In this work we show how to apply the atomic force microscope to perform stress relaxation and creep compliance measurements on living cells. We also propose a derivation of the mechanical properties of the studied biomaterial using these type of experiments. In this way, the relaxation time, the elastic moduli and the viscosity can be obtained using Zener´s model.

## ACKNOWLEDGEMENTS

SMF, MdMV and JLTH thank the ETORTEK programme for financial support.

LITERATURE

Alcaraz, J., Buscemi, L., Grabulosa, M., Trepat, X., Fabry, B., Farre, R., Navajas, D., 2003. Microrheology of human lung epithelial cells measured by atomic force microscopy. Biophysical Journal 84, 2071-2079.

Bausch, A. R., Möller, W., Sackmann, E., 1999. Measurement of local viscoelasticity and foces in living cells by magnetic tweezers. Biophysical Journal 76, 573-579.

Bausch, A. R., Zeimann, F., Boulbitch, A. A., Jacobson K., Sackmann, E., 1998. Local measurements of viscoelastic parameters of adherent cell surfaces by magnetic bead microrheometry. Biophysical Journal 75, 2038-2049.

Bremmell, K.E., Evans, A., Prestidge, C. A., 2006. Deformation and nano-rheology of red blood cells. An AFM investigation. Colloids and Surfaces B: Biointerfaces 50, 43-48.

Butcher, D. T., Alliston, T., Weaver, V. M., 2009. A tense situation. Forcing tumour progression. Nature Reviews Cancer 9, 108-122

Dimitriadis E. K., Horkay F., Maresca J., Kachar, B., Chadwick R. S., 2002. Determination of elastic moduli of thin layers of soft material using the atomic force microscope. Biophysical Journal 82, 2798-2810.

Hochmuth, R.M., 2000. Micropipette aspiration of living cells. Journal of Biomechanics 33, 15-22.

Karcher, H., Lammerding, J., Huang, H., Lee, R. T., Kamm, R. D., Kaazempur-Mofrad, M. R., 2003. A three-dimensional viscoelastic model for cell deformation with experimentl verification. Biophysical Journal, 85, 3336-3349.

Kuznetsova, T. G., Starodubtseva, M. N., Yegorenkov, N. I., Chizhik, S. A., Zhdanov, R. I., 2007. Atomic force microscopy probing of cell elasticity. Micron 38, 824-833.

Leporatti, S., Vergara, D., Zacheo, A., Vergaro, V., Maruccio, G., Cingolani, R., Rinaldi, R., 2009. Cytomechanical and topological investigation of MCF-7 cells by scanning force microscopy. Nanotechnology 20, 055103 (6pp)

Li, Q.S., Lee, G. Y. H., Ong, C. N., Lim, C. T., 2008. AFM indentation study of breast cancer cells. Biochemical Biophysical Research Communications 374, 609-613.

Mathur, A. B., Collinsworth, A. M., Reichert, W. M., Kraus, W. E., Truskey, G. A., 2001. Endothelial, cardiac muscle and skeletal muscle exhibit different viscous and elastic properties as determined by atomic force microscopy. Journal of Biomechanics 34, 1545-1553.

Mahaffy, R. E., Park, S., Gerde, E., Käs, J., Shih, C. K., 2004. Quantitative analysis of the viscoelastic properties of thin regions of fibroblasts using atomic force microscopy. Biophysical Journal 86, 1777-1793.

Melzak, K., Moreno-Flores, S., Yu, K., Kizhakkedathu, J., Toca-Herrera, J. L. *In press.* Rationalized approach to the determination of contact point in force-distance curves: Application to polymer brushes in salt solutions and in water. Microscopy Research and Technique.

Moreno-Flores, S., Benitez, R., Vivanco, M. dM., Toca-Herrera, J. L., 2010. Stress relaxation microscopy: Imaging local stress in cells. Journal of Biomechanics 43, 349-354.

Radmacher, M., Fritz, M., Kacher, C. M., Cleveland, J. P., Hansma, P. K., 1996. Measuring the viscoelastic properties of human platelets with the atomic force microscope. Biophysical Journal 70, 556-567.

Riande, E., Diaz-Calleja, R., Prolongo, M. G., Masegosa, R. M., Salom, C., 2000. Polymer Viscoelasticity. Stress and Strain in Practice. Marcel Decker, New York, p. 879.

Sneddon, I. A., 1965. The relation between load and penetration in the axisymmetric Boussinesq problem for a punch of arbitrary profile. International Journal of Engineering Science 3, 47-57

Thoumine, O., Ott, A., 1997. Time scale dependent viscoelastic and contractile regimes in fibroblasts probed by microplate manipulation. Journal of Cell Science 110, 2109-2116.

Valberg, P. A., Albertini, D. F., 1985. Cytoplasmic motions, rheology, and structure probed by a novel magnetid particle method. Journal of Cell Biology 101, 130-140.

Valberg P. A., Feldman, H. A., 1987. Magnetic particle motions within living cells. Measurement of cytoplasmic viscosity and motile assay. Biophysical Journal 52, 551-569.

Yamada, S., Wirtz, D., Kuo, S.C., 2000. Mechanics of living cells measured by laser tracking microrheology. Biophysical Journal 78, 1736-1737.

Zhao, M., Srinivasan, C., Burgess D. J., Huey, B.D., 2006. Rate and depth dependent nanomechanical behaviour of individual living Chinese Hamster ovary cells probed by atomic force microscopy. Journal of Materials Research 21, 1906-1912.

APPENDIX A. CALCULATION OF ELASTIC MODULI AND VISCOSITIES FROM STRESS RELAXATION AND CREEP EXPERIMENTS (ZENER'S MODEL)

Figure 3 shows the viscoelastic model used to represent the cell's behaviour. In this model the stress can be split into the sum of three terms

$$\sigma = \sigma_0 + \sigma_1 + \sigma_2 \tag{1a}$$

And we have the following equations

$$\begin{aligned} &\sigma_0 = E_0 \varepsilon \\ &\dot{\varepsilon} = \frac{\dot{\sigma}_i}{E_i} + \frac{\sigma_i}{\eta_i}, \text{ for } i\text{=1, 2} \end{aligned} \tag{2a}$$

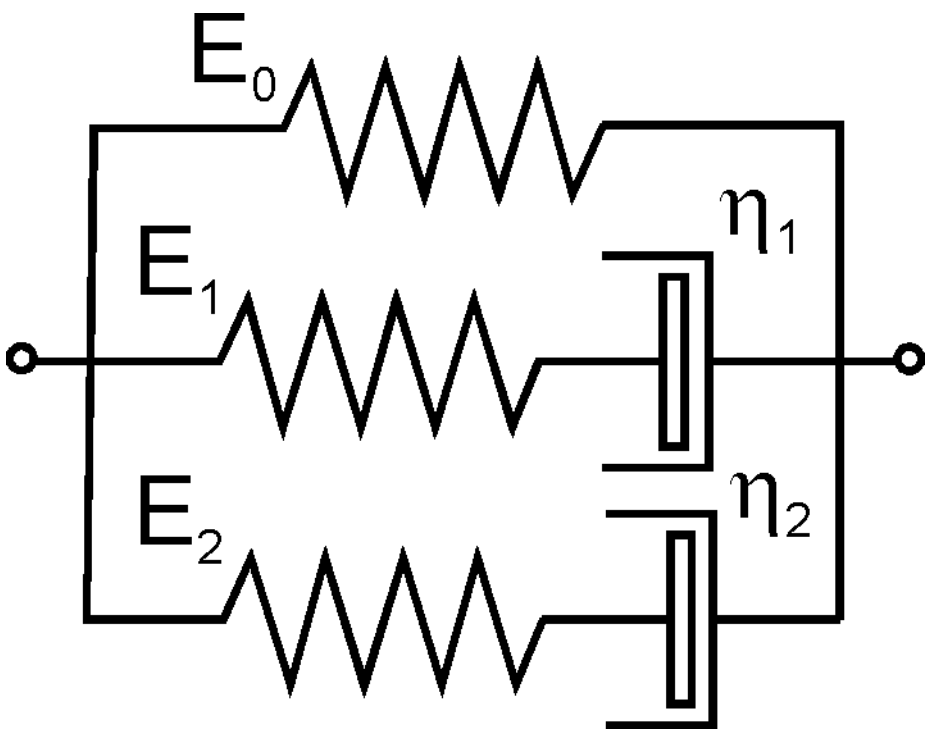

**Figure 3.** Zener's model that accounts in our case for cell behaviour. One elastic and two viscoelastic Maxwell elements arranged in parallel (Riande et al, 2000).

Differentiating twice equation (2a), and taking into account equation (1a), we obtain the following three equations

$$\sigma = E_0 \varepsilon + \sigma_1 + \sigma_2 \tag{3a}$$

$$\dot{\sigma} = \dot{\varepsilon}\left(E_0 + E_1 + E_2\right) - \frac{E_1}{\eta_1}\sigma_1 - \frac{E_2}{\eta_2}\sigma_2 \tag{4a}$$

$$\ddot{\sigma} = \ddot{\varepsilon}\left(E_0 + E_1 + E_2\right) - \dot{\varepsilon}\left(\frac{E_1^2}{\eta_1} + \frac{E_2^2}{\eta_2}\right) + \frac{E_1^2}{\eta_1^2}\sigma_1 + \frac{E_2^2}{\eta_2^2}\sigma_2 \tag{5a}$$

Multiplying equation (3a) by $\frac{E_1 E_2}{\eta_1 \eta_2}$ and equation (4a) by $\left(\frac{E_1}{\eta_1}+\frac{E_2}{\eta_2}\right)$ and after adding these equations to (5a) we get

$$\ddot{\sigma} + A\dot{\sigma} + B\sigma = r_2\ddot{\varepsilon} + r_1\dot{\varepsilon} + r_0\varepsilon \tag{6a}$$

where $A = \frac{E_1}{\eta_1} + \frac{E_2}{\eta_2}$, $B = \frac{E_1 E_2}{\eta_1 \eta_2}$, $r_0 = \frac{E_0 E_1 E_2}{\eta_1 \eta_2}$, $r_1 = \frac{E_1}{\eta_1}\left(E_0 + E_2\right) + \frac{E_2}{\eta_2}\left(E_0 + E_1\right)$ and $r_2 = E_0 + E_1 + E_2$.

*Experiments at constant strain: stress relaxation.* Let us consider the particular case of a constant strain $\varepsilon(t) = \varepsilon_0$, then equation (6a) reduces to

$$\ddot{\sigma} + A\dot{\sigma} + B\sigma = r_0\varepsilon_0 \tag{7a}$$

The general solution to equation (7a) has the form $\sigma(t) = \sigma_p(t) + \sigma_0(t)$ where $\sigma_0(t)$ is the general solution of the homogeneous equation $\ddot{\sigma} + A\dot{\sigma} + B\sigma = 0$ and $\sigma_p(t)$ is a particular solution of (7a). Since the right hand of equation (7a) is constant we may find a constant particular solution $\sigma_p(t) = E_0\varepsilon_0$. On the other hand the general solution of the homogeneous equation takes the form $\sigma_0(t) = A_1 e^{-t/\tau_1} + A_2 e^{-t/\tau_2}$, being $\tau_i = \frac{\eta_i}{E_i}$ for *i*= 1, 2. Therefore the general solution of equation (7a) is

$$\sigma(t) = E_0\varepsilon_0 + A_1 e^{-t/\tau_1} + A_2 e^{-t/\tau_2} \tag{8a}$$

*Experiments at constant height: creep.* Considering now a constant stress $\sigma(t) = \sigma_0$ equation (6a) takes the form

$$r_2\ddot{\varepsilon} + r_1\dot{\varepsilon} + r_0\varepsilon = B\sigma_0 \tag{9a}$$

The particular solution to this equation is given by the constant function $\varepsilon_p(t)=\frac{B\sigma_0}{r_0}=\frac{\sigma_0}{E_0}$. The general solution to the homogeneous equation $r_2\ddot{\varepsilon}+r_1\dot{\varepsilon}+r_0\varepsilon=0$ is given by $\varepsilon_0(t)=C_1e^{x_1t}+C_2e^{x_2t}$; $x_1$ and $x_2$ are the roots of the characteristic polynomial $r_2x^2+r_1x+r_2$ which are given by

$$x_1=\frac{-r_1+\sqrt{r_1^2-4r_0r_2}}{2r_2},\quad x_2=\frac{-r_1-\sqrt{r_1^2-4r_0r_2}}{2r_2} \tag{10a}$$

And both are real and negative, because $r_1^2-4r_0r_1=\left(\frac{E_1}{\eta_1}(E_0+E_2)-\frac{E_2}{\eta_2}(E_0+E_1)\right)^2+\frac{4E_1^2E_2^2}{\eta_1\eta_2}>0$. Thus the general solution to equation (9) is

$$\varepsilon(t)=\frac{\sigma_0}{E_0}+C_1e^{x_1t}+C_2e^{x_2t} \tag{11a}$$

*Obtaining parameters*. We assume we have experimentally obtained two signals that follow Zener's model. Then we have

$$\sigma(t)=A_0+A_1e^{-t/\tau_1}+A_2e^{-t/\tau_2} \tag{12a}$$

$$\varepsilon(t)=C_0+C_1e^{x_1t}+C_2e^{x_2t} \tag{13a}$$

To obtain the coefficients $E_0$, $E_1$, $E_2$, $\eta_1$, $\eta_2$ and $\eta_3$ from the experimental coefficients $A_0$, $\tau_1$, $\tau_2$, $C_0$, $x_1$ and $x_2$, we assume we know $\sigma_0$ and $\varepsilon_0$. $E_0$ is thus easily obtained as

$$E_0=\frac{A_0}{\varepsilon_0} \tag{14a}$$

We can in turn get the value of $r_0$ from its definition and $E_0$, $r_0=\frac{E_0E_1E_2}{\eta_1\eta_2}=\frac{E_0}{\tau_1\tau_2}$.

Knowing $r_0$, we can obtain $r_1$ and $r_2$ in terms of $x_1$, $x_2$ and $r_0$ by multiplying the expressions of $x_1$ and $x_2$, $x1x2 = \frac{r_1^2 - \left(\sqrt{r_1^2 - 4r_0 r_2}\right)^2}{4r_2^2} = \frac{4r_0 r_2}{4r_2^2} = \frac{r_0}{r_2}$. We thus obtain $r_2$ as

$$r_2 = \frac{r_0}{x_1 x_2} \tag{15a}$$

and $r_1$ as

$$r_1 = -r_2 x_1 - \frac{r_0}{x_1} = -r_0\left(\frac{1}{x_1} + \frac{1}{x_2}\right) \tag{16a}$$

We can then rewrite the expressions of $r_2$ and $r_1$ in terms of $E_0$, $\tau_1$, $\tau_2$, $r_1$ and $r_2$ as follows

$$\begin{aligned} &E_1 + E_2 = r_2 - E_0 \\ &\frac{E_1}{\tau_2} + \frac{E_2}{\tau_1} = r_1 - E_0\left(\frac{1}{\tau_1} + \frac{1}{\tau_2}\right) \end{aligned} \tag{17a}$$

Equations (17a) are a system of two linear equations with two unknowns ($E_1$ and $E_2$). The solution gives

$$\begin{aligned} &E_1 = r_2 - \left[E_0 + \frac{1}{\left(\frac{1}{\tau_1} - \frac{1}{\tau_2}\right)} \cdot \left(r_1 - \frac{r_2}{\tau_2} - \frac{E_0}{\tau_1}\right)\right] \\ &E_2 = \frac{1}{\left(\frac{1}{\tau_1} - \frac{1}{\tau_2}\right)} \cdot \left(r_1 - \frac{r_2}{\tau_2} - \frac{E_0}{\tau_1}\right) \end{aligned} \tag{18a}$$

which turns to

$$E_1 = \frac{A_0}{\varepsilon_0}\left[\frac{1}{\left(1-\frac{\tau_1}{\tau_2}\right)}\cdot\left(1+\frac{1}{x_1\tau_2}+\frac{1}{x_2\tau_2}+\frac{1}{x_1x_2\tau_2^2}\right)+\frac{1}{x_1x_2\tau_1\tau_2}-1\right]$$
$$E_2 = \frac{A_0}{\varepsilon_0\left(1-\frac{\tau_1}{\tau_2}\right)}\cdot\left(-\frac{1}{x_1x_2\tau_2^2}-\frac{1}{x_1\tau_2}-\frac{1}{x_2\tau_2}-1\right) \quad (19a)$$

as a function of the experimental parameters. Once $E_1$ and $E_2$ are known, it is possible to calculate the viscosities $\eta_1$ and $\eta_2$ by substitution into their respective expressions

$$\eta_1 = E_1\tau_1 = \frac{A_0\tau_1}{\varepsilon_0}\left[\frac{1}{\left(1-\frac{\tau_1}{\tau_2}\right)}\cdot\left(1+\frac{1}{x_1\tau_2}+\frac{1}{x_2\tau_2}+\frac{1}{x_1x_2\tau_2^2}\right)+\frac{1}{x_1x_2\tau_1\tau_2}-1\right]$$
$$\eta_2 = E_2\tau_2 = \frac{A_0\tau_2}{\varepsilon_0\left(1-\frac{\tau_1}{\tau_2}\right)}\cdot\left(-\frac{1}{x_1x_2\tau_2^2}-\frac{1}{x_1\tau_2}-\frac{1}{x_2\tau_2}-1\right) \quad (20a)$$